\documentclass{article}
\usepackage{spconf,amsmath,graphicx} 
\usepackage{multirow} 
\usepackage{amssymb}
\usepackage[numbers,sort&compress]{natbib}
\usepackage{enumitem}
\usepackage[dvipsnames]{xcolor}

\title{REDAT: Accent-Invariant Representation for End-to-End ASR by\\ Domain Adversarial Training with Relabeling}
\name{
\begin{tabular}{@{}c@{}}
    Hu Hu\sthanks{This work was performed while the author was an intern at Amazon.}, Xuesong Yang$^\dagger$, Zeynab Raeesy$^\dagger$, Jinxi Guo$^\dagger$, Gokce Keskin$^\dagger$,\\
    Harish Arsikere$^\dagger$, Ariya Rastrow$^\dagger$, Andreas Stolcke$^\dagger$, Roland Maas$^\dagger$
\end{tabular}
}
\address{$^\ast$Georgia Institute of Technology, Atlanta, USA \qquad $^\dagger$Amazon Alexa}
\begin{document}
%
\maketitle
\begin{abstract} 
    Accents mismatching is a critical problem for end-to-end ASR\@. This paper aims to address this problem by building an accent-robust RNN-T system with
    domain adversarial training (DAT).
    We unveil the magic behind DAT and provide, for the first time, a theoretical guarantee that DAT learns accent-invariant representations. We also prove that performing the gradient reversal in DAT is equivalent to minimizing the Jensen-Shannon divergence between domain output distributions.
    Motivated by the proof of equivalence, we introduce \emph{reDAT}, a novel technique based on DAT, which
    relabels data using either unsupervised clustering or
    soft labels. 
    Experiments on 23K hours of multi-accent data show that DAT achieves competitive
    results over accent-specific baselines on both native and non-native English accents but up to 13\% relative WER reduction on unseen accents; our \emph{reDAT} yields further improvements over DAT by 3\% and 8\% relatively on non-native accents of American and British English. 
\end{abstract}
\section{Introduction} 
\label{sec:intro}

Recent application of recurrent neural network transducers (RNN-T) has achieved significant progress in the area 
of online streaming end-to-end automatic speech recognition (ASR)~\cite{rnnt-google1, rnnt-google2, rnnt-ms, rnnt-jinxi}.
However, building an accent-robust system remains a big challenge. 
Accents represent systematic variations within a language as a function of geographical region (e.g. British versus American English), social group, or other factors such as nativeness of speakers. Accents occur in many gradations and commercial speech applications typically only model varieties associated with major countries.
For example in real-world smart speaker devices, users set up their language preferences regardless of whether
they are native speakers or not; thus ASR systems trained mainly on only native speech
risk degradation when faced with non-native speech.

Accent-robust ASR systems aim to mitigate the negative effects of non-native speech.
A straightforward exploration is to build an
accent-specific system where accent information, such as i-vectors, accent IDs, or accent
embeddings, are explicitly fed into the neural networks along with acoustic features~\cite{e2e-emb-ivector,accent-multi,accent-bengio,accent-xuesong,accent-google,accent-ms}. 
These
approaches typically either adapt a unified model with accent-specific data, or build a separate
decoder for each accent. Accent-specific models perform well on the test sets with consistent accents,
but they do not generalize well to unseen accents. Accent-invariant systems~\cite{pooling1, pooling2}, 
alternatively, build a universal model to learn accent-invariant features that are expected to generalize well to new accents.
For example, simply pooling data across all accents during training brings in additional variations
so that the models are capable of learning accent-invariant information;
adversarial training methods~\cite{aipnet,dat-accent} also help to achieve the same goal
through the gradient reversal.

We aim to advance accent-invariant modeling with \mbox{RNN-T} based on the domain adversarial training (DAT)~\cite{dat}. DAT is expected to learn accent-invariant features by reversing gradients propagated from the accent classifier. Our experiments demonstrate DAT can achieve competitive performance on native, non-native, and unseen accents.
This paper makes the following novel contributions: 
\vspace*{-0.6em}
\begin{itemize}[itemsep=1mm, parsep=0pt]
    \item We lay out the theory behind DAT and we provide, for the first time, a theoretical guarantee that DAT learns accent-invariant representations.
    \item We also prove that performing the gradient reversal in DAT is equivalent to minimizing the Jensen-Shannon divergence between output distributions from different domain classes.
    \item Motivated by the proof of equivalence, we introduce \mbox{\emph{reDAT}}, a novel technique based on DAT, which refines accent classes with either unsupervised clustering or soft labels. Our \emph{reDAT} yields significant improvements over strong baselines on non-native and unseen accents without sacrifice of native accents performance.
\end{itemize}

\section{DAT for Accented Speech Recognition} 
\label{sec:dat}

\begin{figure}[t] 
    \centering 
    \includegraphics[width=\linewidth]{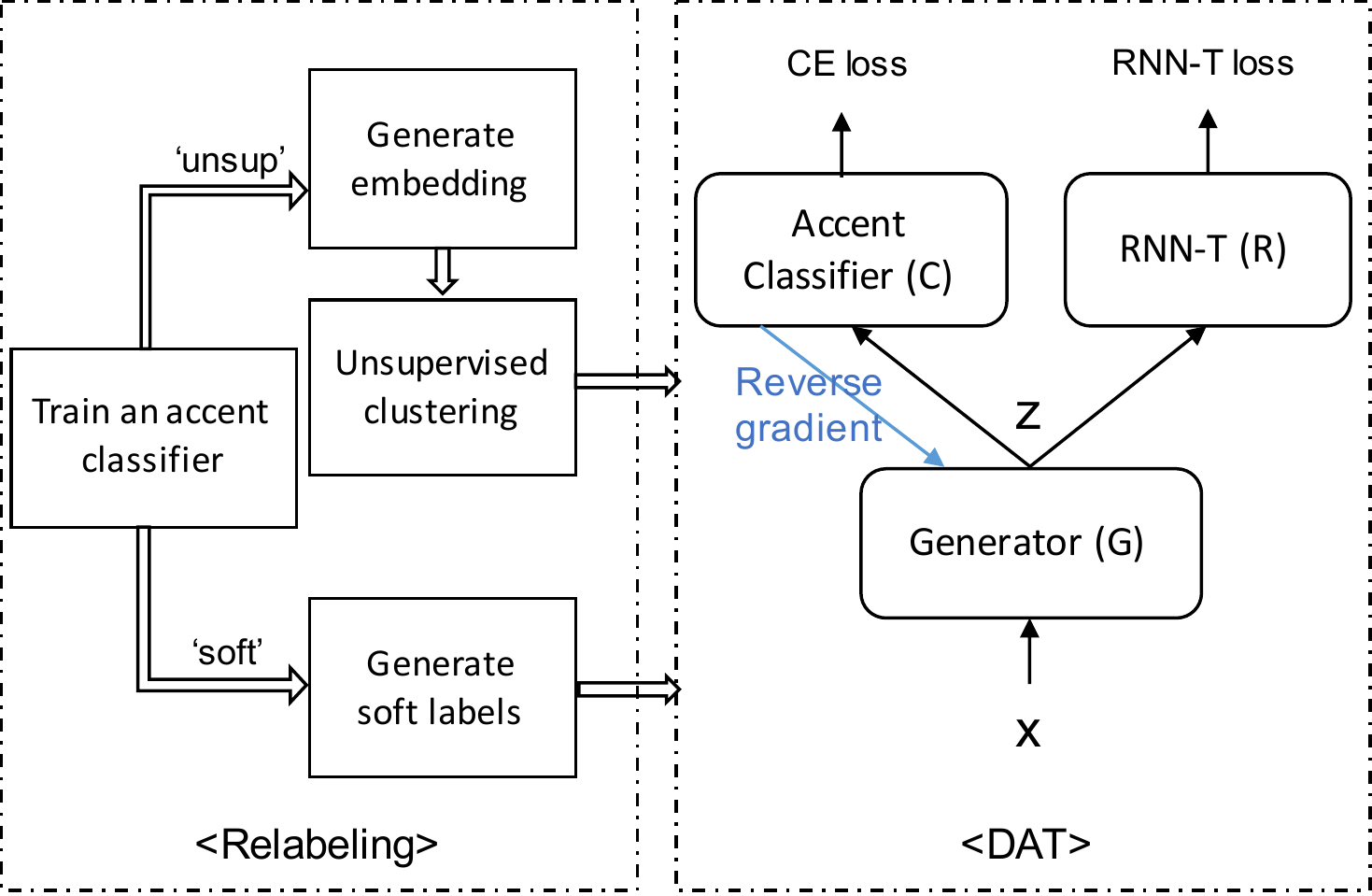}
    \caption{REDAT framework by relabeling with either unsupervised clustering (`unsup') or soft labels (`soft').} 
    \label{fig:framework}
\end{figure}

Domain adversarial training (DAT) has been widely applied to robust ASR systems under multiple conditions including speakers~\cite{dat-speaker}, noises~\cite{dat-noise}, accents~\cite{dat-accent}, and languages~\cite{dat-multilingual}. Our proposed DAT training framework consists of an accent-invariant feature generator $G$, English accent classifier $C$, and RNN-T model $R$ (see Figure~\ref{fig:framework}).
LSTM layers are used for the feature generator and accent classifier. Our RNN-T model includes encoder, decoder, and joint networks. During training,
negative gradients~(\textcolor{RoyalBlue}{blue arrow}) are back-propagated to the generator from the accent classifier so that its ability of distinguishing accents embedded in the generator outputs is minimized. In other words, the output $z$ from the generator $G$ is expected to embed accent-invariant representations.

We denote losses of $R$ and $C$ as $\mathcal{L}_G$ and $\mathcal{L}_C$, the weight matrices of $G$, $C$, $R$ as $\theta_G$, $\theta_C$, $\theta_R$. Each weight is updated by the following gradient descent rules,
\small
\begin{align*}
    \theta_G &\leftarrow \theta_G - \alpha \left(\frac{\partial \mathcal{L}_{R}}{\partial \theta_G} - \lambda \frac{\partial \mathcal{L}_{C}}{\partial \theta_G}\right), \nonumber \\
    \theta_C &\leftarrow \theta_C - \alpha \frac{\partial \mathcal{L}_{C}}{\theta_C}, \nonumber\\
    \theta_{R} &\leftarrow \theta_{R} - \alpha \frac{\partial \mathcal{L}_{R}}{\theta_{R}}, \nonumber 
\end{align*}
\normalsize
where $\alpha$ is the learning rate and $\lambda$ is the scale of $\mathcal{L}_C$ gradients.
When making forward inference, we freeze $G$ and $R$.

\subsection{Theoretical Guarantees of DAT for Accent-Invariance}
DAT is capable of learning expressive domain-invariant features in practice but we have little theories to explain the magic under the hood. Ganin et al~\cite{dat} tried to explain DAT in a two-domain scenario, and their theory was established on the notion of H-divergence where the distance between source-target domains is minimized. H-divergence theory itself is rather limited in understanding domain adaptation problems. The inherent principle of DAT is to minimize the Jensen-Shannon divergence (JSD) of the source-target distributions~\cite{gan}, which indicates JSD helps to overcome the limitations of H-divergence and provides an alternative foundation for a better understanding. We extend the theory behind GANs~\cite{gan}, and prove that performing gradient reversal is equivalent to minimizing JSD among multiple domain distributions. Our findings could generalize to any domain mismatch problems. We focus on accents in this paper.

The output distribution of the $i$-th accent from $G$ is denoted as $P_{G_i}$ where $i\in[1, N]$ and $N$ is the number of observed accents.
Given that $G$ is fixed during training, we could find the optimal $C^\ast$ by minimizing their cross-entropy (CE) or equivalently maximizing the log-likelihood as,
\small
\begin{align} 
    C^\ast = \mathop{\arg\max}_{\theta_C} \sum_i^N E_{z \sim P_{Gi}(z)} \log C_i (z).
    \label{equ:optimal_c_expression} 
\end{align}
\normalsize
We use \emph{softmax} to normalize the final outputs such that the probability $C_i(z)$ of an input utterance belonging to the $i$-th accent must satisfy,
\small
\begin{align}
    \sum_{i}^N C_i (z) = 1,\ \ \ \ \ \ 0 < C_i (z) < 1.
    \label{equ:constraints} 
\end{align}
\normalsize
Eq~\eqref{equ:optimal_c_expression}~\eqref{equ:constraints} indicate $C^\ast$ is convex such that it must have a global maxima since the 2nd-order derivative of $C_i(z)$ is always negative. We can then find the solution by linear programming,
\small
\begin{align} 
    C_i^\ast(z) = \frac{P_{G_i}}{\sum_i^N P_{G_i}}.
    \label{equ:optimal_c} 
\end{align}
\normalsize
Our solution is similar to GANs~\cite{gan} but extends to multiple variables. The generator $G$ connects two tasks so that two different losses are accessed.
$\mathcal{L}_{C}$ is the CE loss of $C$, and it propagates its
negative gradients to $G$. If we only consider the effect of $\mathcal{L}_C$ on $G$, we have the optimal
$G^\ast$ as, 
\small
\begin{align} 
    G^\ast = \mathop{\arg\min}_{\theta_G}  \left(\mathop{\arg\max}_{\theta_C}
    \sum^N_i E_{x \sim P_{data}(x)} \log C_i \left(G(x)\right)\right).
    \label{equ:optimal_g} 
\end{align}
\normalsize
When updating parameters of $G^\ast$, $C$ is fixed. We can find the solution of $C_i^\ast
(z)$ by plugging Eq~\eqref{equ:optimal_c} into Eq~\eqref{equ:optimal_g}. After deduction
and simplification, we have the optimal $G^\ast$ as, 
\small
\begin{align} 
    G^\ast = \mathop{\arg\min}_{\theta_G} \left(-N\log N + \sum_i^N KLD \left(P_{G_i} \parallel \frac{\sum_i^N
    P_{G_i}}{N}\right)\right), \nonumber
\end{align} 
\normalsize
where $KLD$ is the Kullback–Leibler divergence. We can reformulate it equivalently by minimizing the JSD across the distributions of all accents as,
\small
\begin{align} 
    G^\ast = \mathop{\arg\min}_{G} \left(-N\log N + JSD\left(P_{G_1}, P_{G_2}, \ldots, P_{G_N}\right)\right). \nonumber
\end{align}
\normalsize
We conclude the proof that performing gradient reversal is equivalent to minimizing the
Jensen-Shannon divergence between output distributions from different accents. The
global minima is achieved if and only if $P_{G_1}$=$P_{G_2}$=$\ldots$=$P_{G_N}$, which indicates that the embeddings $z$ are accent-invariant.

\section{REDAT: DAT with Relabeling} 
\label{sec:relabel} 
The theoretical proof of the equivalence
between performing gradient reversal and minimizing JSD of output distributions from accents suggests that we could get more invariant training results by predefining more detailed acoustic information, such as a refined accent label for utterances.
Accents boundaries are not well defined in practice and realistic accent-specific data is usually mixed with native and non-native accents. In order to mitigate the drawbacks, we further propose a novel method, \emph{reDAT}, to refine labels of domain classes either by unsupervised clustering or with soft labels.

\subsection{Relabeling with Unsupervised Clustering} 
We relabel utterance accents in a three-phase unsupervised manner (see \emph{unsup} in Figure~\ref{fig:framework}).
An utterance-level accent classifier is trained with original accent labels;
we then extract utterance-level embeddings using this well-trained accent
classifier where distinct accents information is detailed;
lastly, we predict new domain labels for utterances by performing $k$-means clustering on these utterance embeddings.
DAT could directly benefit from the newly generated accent domain labels and improve its generalization ability to non-native and unseen accents.
We specify an optimal number of clusters $k$ larger than the number of accent variants existing on our training data where clear boundaries across $k$ clusters are observed from t-SNE visualization.
We also increase $k$ to capture detailed English accents that are transferred from non-English native languages. There must be extra effort to estimate the ideal number of challenging non-native accents, but it is beyond the scope of this paper.

\subsection{Relabeling with Soft Labels} 
We can also refine accent labels with soft labels in a two-phase process (see \emph{soft} in 
Figure~\ref{fig:framework}). 
An utterance-level accent classifier is trained with original accent labels; we then
generate a soft label for each utterance from this accent classifier. 
DAT is performed based on these newly generated soft labels.
Previous studies found that soft labels correlate with structural relationship among accents~\cite{rkd, nle-rtsl} so that we expect them to encode more detailed accent information. Although one-hot labels are replaced by soft labels, our theoretical 
equivalence of performing gradient reversal and minimizing JSD still holds, but this JSD is accessed between each utterance distribution. 
When one-hot labels are replaced by soft labels, the expression of $C^\ast$ in
Eq~\eqref{equ:optimal_c_expression} is reformulated as, 
\small
\begin{align} 
    C^\ast = \mathop{\arg\max}_{\theta_C}  E_{z \sim P_{G}(z)} \sum_i^N l_i(x) \log C_i(z), \nonumber
    \label{equ:optimal_c_expression_soft} 
\end{align} 
\normalsize
where $l_i(x)$ is the scalar soft label of an input utterance $x$ predicted by the accent classifier $l_i$.
Then $G^\ast$ is derived as,
\small
\begin{align} 
    G^\ast = \mathop{\arg\min}_{\theta_G} \ \left(-NlogN + JSD\left(l(x_1)\cdot P_G, \ldots, l(x_N)\cdot P_G\right)\right), \nonumber
\end{align} 
\normalsize
where $l(x_i)\cdot P_G$ is a distribution depending on the input
$x$, which can be regarded as the linear combination of different accent distributions. Thus, by
using soft labels for gradient reversal, we replace minimizing JSD between accent distributions with doing the same
between utterance distributions. 

\section{Experiments} 
\label{sec:exp}

\begin{table*}[htbp] 
    \centering 
    \caption{Normalized WERs$^1$ on 23K
    hours of en-X data. \emph{AS} or \emph{AI} denotes an accent-specific or accent-invariant
    model; \emph{native} or \emph{non-native} denotes native or non-native speakers on test
    sets; \emph{unsup8} or \emph{unsup20} denotes \emph{reDAT} with 8 or 20
    unsupervised clusters; \emph{soft} denotes \emph{reDAT} with soft labels.} 
    \label{tab:all} 
    \vspace{0.3cm} 
    \begin{tabular}{l|c|c|c|c|c|c|c|c} 
        \hline 
        \hline
        \multirow{2}{*}{Approach} & \multirow{2}{*}{AS/AI} & \multicolumn{3}{c|}{en-US \%} & \multicolumn{3}{c|}{en-GB \%} & \multirow{2}{*}{\begin{tabular}[c]{@{}c@{}}en-AU \%\\
        (unseen) \end{tabular}} \\ \cline{3-8} &                        
        & \ \ \ native\ \ \  & non-native &\ \ \ \ avg.\ \ \ \         &\ \ \ native\ \ \   & non-native  &\ \ \ \ avg.\ \ \ \  &  \\
        \hline
\hline 
M0: Data pooling              & AI                     & 1.000 & 1.472    & 1.027          & 1.315   &  1.574    & 1.315  & 1.393\\
M1: AIPNet-s                  & AI                     & 0.997 & 1.425    & 1.023          & 1.330   & 1.543     & 1.332  & 1.412\\
\hline 
M2: One-hot embeddings         & AS                     & 0.981 & 1.528    & 1.010 & 1.284   & 1.540     & 1.284  & 1.574\\
M3: Linear embeddings          & AS                     & 0.991 & 1.442    & 1.017 & 1.284   & 1.534     & 1.282  & 1.569\\
\hline 
M4: DAT                       & AI                     & 0.985 & 1.448    & 1.012 & 1.293  & 1.567     & 1.294  & 1.373\\
M5: reDAT-unsup8        & AI                     & \textbf{0.969} & 1.472 &\textbf{0.996} & \textbf{1.270} & 1.465 & \textbf{1.266} & \textbf{1.359}\\
M6: reDAT-unsup20       & AI                     & 0.980 & 1.470    & 1.006 & 1.282   & 1.492     & 1.280  & 1.361\\
M7: reDAT-soft          & AI                     & 0.973 & \textbf{1.409} & 0.997   & 1.309   & \textbf{1.440}     & 1.307  & 1.388  \\
        \hline 
        \hline 
    \end{tabular}
\end{table*} 

\subsection{Experimental Setup} 
For our experiments we used de-identified human labeled speech data (23K hours) from voice controlled far-field and close-talk devices. This data set consists of English recordings from 3 different regions, including 13K hours of
en-US data, 6K hours of en-GB data, and 4K hours of en-IN (Indian English) data. Each utterance in the en-US and en-GB test sets has a label that characterizes the speaker as native or non-native.
Most of the recordings (over 90\%) are from native speakers.
In addition, to evaluate generalization, we use extra en-AU (Australian English) data as an unseen test set. Since the en-IN data lacks nativeness labels and is smaller in size,
we only evaluate on en-US, en-GB, and en-AU test sets. 

All experiments use 64-dimensional log-Mel features, computed over 25ms windows with 10ms hop
length. Each feature vector is stacked with 2 frames to the left and down-sampled to a 30ms frame
rate. All experiments are performed with an RNN-T model. The baseline RNN-T model consists of an
encoder, a prediction network, and a joint network. The encoder consists of 5 LSTM layers with the
hidden dimension of 1024, whereas the prediction network consists of 2 LSTM layers with the hidden
dimension of 1024 and the embedding size of 512. 
We adopt a simple addition strategy in the joint network to combine outputs from encoder and prediction networks to limit memory and computation.
The softmax layer consists
of 10K output units and is trained to predict word-piece tokens, which are generated using the byte
pair encoding algorithm~\cite{bpe}. To apply the \emph{reDAT} framework to the RNN-T model, the first two LSTM
encoder layers of RNN-T serve as the generator, whose outputs are fed into a domain classifier as well as
into the remaining parts of the RNN-T encoder. The accent classifier consists of 2 LSTM layers with the
hidden dimension of 1024 and predicts three accent classes, i.e.\ en-US, en-GB, and en-IN. 

All models are trained using the Adam optimizer~\cite{adam}, with a learning rate schedule including
an initial linear warm-up phase, a constant phase, and an exponential decay phase~\cite{rnnt-jinxi}. All the baseline models and proposed methods use the same training
strategy. Specifically, the learning rates for the constant phase and end of the exponential decay
phase are $5e-4$ and $1e-5$, respectively. During the training stage, the acoustic training data is
augmented with the SpecAugment~\cite{specaug} to improve the robustness. 

\subsection{Baselines} 
We investigate the state-of-the-art multi-accent ASR systems and choose two accent-invariant approaches (M0,M1) and two accent-specific (M2,M3) approaches as our baselines (see Table~\ref{tab:all}).
Data pooling (M0) combines data of all accents together and trains a unified model.
Accent-specific systems utilize external accent information~\cite{accent-google} and append embeddings to the outputs of each layer in the \mbox{RNN-T} model. Specifically,
one-hot embeddings (M2) directly uses one-hot accent labels whereas linear embeddings (M3) applies a matrix to map one-hot labels into linear embedding vectors.
Accent-specific systems require consistent accents in both training and evaluation phases so that they are vulnerable to unseen accents not existing on the training data.
\mbox{AIPNet}~\cite{aipnet} introduces an extra accent-invariant GAN and decoder layer for pre-training and jointly trains ASR model and invariant feature generator altogether. We simplify it as \mbox{AIPNet-s} by replacing accent-specific GAN with one-hot labels in order to provide a stronger baseline. \mbox{AIPNet-s} (M1) directly uses oracle accent embeddings so that it is expected to contribute an upper-bound performance of AIPNet.

\subsection{Experimental Results on 23K Hours of en-X Data} 
The experimental results of the normalized
word error rates (WERs)\footnote{Normalized WER of a control model is calculated as the WER percentage over the reference. For example, \emph{Data Pooling} is chosen as the reference so that its WER is 1.000, and \emph{DAT}, as a control, is 0.985.} are shown in Table~\ref{tab:all} where the performance of our baseline system is below 10\% WER absolute. At first, when comparing the results
on native and non-native speakers, we can see that although we may achieve good ASR
performance on native speakers, there is still a big performance gap between native speakers and
non-native speakers. Results of all baselines (M0,M1,M2,M3) are described in Table~\ref{tab:all}.
As for accent-invariant baselines, \mbox{AIPNet-s} can bring gains on non-native data over data
pooling. As for accent-specific approaches, i.e.,\@ one-hot embeddings and linear embeddings,
they show better performance than data pooling on native data, but do not generalize well to the unseen accent test set.
That is because these two models do not know the accent label for the en-AU test set, and they have to choose an accent-specific model (e.g.\ trained on en-US) to make evaluations, even though the accents are mismatched.

Experimental results of DAT and our proposed \emph{reDAT} are shown in the last four rows of Table~\ref{tab:all}. 
When compared to \emph{AI} and \emph{AS} baselines,
DAT achieves competitive WERs on both native and non-native accents but up to 13\% relative reduction on unseen accents; the best performance of \emph{reDAT} with 8 unsupervised clusters shows relative WER reductions of
2\% to 4\% over the data pooling baseline and 2\% over DAT, respectively. When increased to 20 unsupervised clusters, we observe a WER degradation over 8 clusters. On non-native accents,
our \emph{reDAT} with soft labels achieves significant improvements over DAT by 3\% on en-US and 8\% on en-GB, and over the best \emph{AI} and \emph{AS} baselines by 1\% on en-US and 6\% on en-GB\@. On native and unseen accents, we observe that
\emph{reDAT} with soft labels has very competitive results 
over original DAT.

\section{Conclusion} 
\label{sec:con} 
This paper suggests a feasible solution to address accents mismatching problems for
end-to-end RNN-T ASR using DAT\@. We demonstrate that DAT could achieve competitive WERs over accent-specific baselines on both native and non-native English accents but significantly better WERs on unseen accents. We provide, for the first time, a theoretical guarantee that DAT extracts accent-invariant representations that generalize well across accents, and also prove that performing gradient reversal in DAT is equivalent to minimizing JSD between domain distributions. The proof of equivalence further motivates to introduce a novel method \emph{reDAT} that yields relative WERs over DAT on non-native accents by a large margin.


\newpage 
\newpage

\small 
\bibliographystyle{IEEEbib} 
\bibliography{refs}

\end{document}